\documentclass[a4paper, 10 pt]{ieeeconf} 

\usepackage[utf8]{inputenc}
\usepackage[T1]{fontenc}
\usepackage{graphicx}
\usepackage{mathtools}
\usepackage{amsfonts}
\usepackage{amssymb}
\usepackage{caption}
\usepackage{subcaption}
\usepackage{color}
\usepackage{textcomp}
\usepackage{url}
\usepackage{booktabs}
\usepackage{array}

\begin{document}








\title{\LARGE \bf Dynamics of collective action to conserve a large common-pool resource
}

\author{\authorblockN{David Andersson\authorrefmark{1}\authorrefmark{2}, Sigrid Bratsberg\authorrefmark{2}, Andrew K. Ringsmuth\authorrefmark{3}\authorrefmark{5}\authorrefmark{4}\authorrefmark{6}\authorrefmark{7}, Astrid S.\ de Wijn \thanks{Corresponding author: Astrid de Wijn (astrid.dewijn@ntnu.no)}\authorrefmark{2}\authorrefmark{1}}\\
\authorblockA{\authorrefmark{1}Chemical Physics Division, Department of Physics, Stockholm University, Sweden\\}
\authorblockA{\authorrefmark{2}Department of Mechanical and Industrial Engineering, Norwegian University of Science and Technology, Norway\\}
\authorblockA{\authorrefmark{3}Complexity Science Hub Vienna, Vienna, Austria\\} 
\authorblockA{\authorrefmark{4}Section for the Science of Complex Systems, Medical University of Vienna, Vienna, Austria\\}
\authorblockA{\authorrefmark{5}Stockholm Resilience Centre, Stockholm University, Stockholm, Sweden\\}\authorblockA{\authorrefmark{6}Current Address: Wegener Center for Climate and Global Change, University of Graz, Graz, Austria\\}\authorblockA{\authorrefmark{7}Current Address: Complexity Science Hub Vienna, Vienna, Austria}}


\maketitle


\begin{abstract}
A pressing challenge for coming decades is sustainable and just management of large-scale common-pool resources including the atmosphere, biodiversity and public services. This poses a difficult collective action problem because such resources may not show signs that usage restraint is needed until tragedy is almost inevitable. To solve this problem, a sufficient level of cooperation with a pro-conservation behavioural norm must be achieved, within the prevailing sociopolitical environment, in time for the action taken to be effective. Here we investigate the transient dynamics of behavioural change in an agent-based model on structured networks that are also exposed to a global external influence.
We find that polarisation emerges naturally, even without bounded confidence, but that for rationally motivated agents, it is temporary.
The speed of convergence to a final consensus is controlled by the rate at which the polarised clusters are dissolved.
This depends strongly on the combination of external influences and the network topology. Both high connectivity and a favourable environment are needed to rapidly obtain final consensus.
\end{abstract}


\section{INTRODUCTION}
Common-pool resources (CPRs) are `non-excludable' (open-access) and `subtractable' (one individual's consumption can preclude another's). Harvesting a CPR for short-term benefit without restraint can make it unavailable through depletion -- a so-called `tragedy of the commons' -- unless harvesters can organise to restrain their harvesting to a sustainable level. 

CPR management has typically been studied in systems where harvesters can perceive the resource level's linear response to their harvesting, allowing them to adapt efforts accordingly (e.g. pastures, groundwater, fisheries). In models, this appears as explicit interdependence between harvesting strategy evolution and resource level \cite{ringsmuth2019, min2018,chen2018,szolnoki2017,dobay2014,tavoni2012}. However, a CPR's linear response may be imperceptible to harvesters if, for example, the resource pool is very large when compared with the harvesting rate. Such a resource may show no obvious reason for restraint, even after extensive harvesting. However, the state of the encompassing social-ecological system  may depend nonlinearly on the resource level, so unrestrained harvesting may push the system to an unexpected critical transition into a new and potentially undesirable state \cite{scheffer2012}. Three prominent examples of such CPRs are unpolluted atmosphere, global biodiversity and public healthcare systems during a pandemic.

Conservation of a large CPR can be considered as a large-scale (potentially global) collective action dilemma \cite{ostrom2011,buck2012}. In such a dilemma, group members must choose whether to contribute to a collective action such as adherence to a pro-conservation behavioural norm which, if successful, will benefit the whole group. Below some threshold rate of cooperation, each individual benefits more from defecting than cooperating, but all individuals are better off if the threshold is reached \cite{heckathorn1996}.

The lack of obvious feedback from a large CPR amplifies the collective action problem because the need for restraint may not be obvious to enough people until tragedy is (almost) inevitable. Solving this problem requires achieving a sufficient level of cooperation with a pro-conservation behavioural norm fast enough for the collective action to successfully conserve the resource. Precisely what this level and rate are depends on the details of a given CPR and how it responds to changes in harvesting behaviour. In all cases, however, the behavioural change process will depend on agents being affected by information received from other agents and from the broader sociopolitical environment. In general, the faster and more complete this process is, the more likely it is that the conservation action will be successful. 

Models of opinion dynamics on social networks have been extensively studied~\cite{bcmreview, perc}. Many are agent-based and assume `bounded confidence': an agent's interaction with another may change its opinion but only if they engage meaningfully. Due to the psychosocial phenomenon of homophily, this is unlikely to happen between people who have very different views. Bounded confidence models~\cite{deffuant,hegselkrause} therefore assume that agents engage only if their opinions differ by less than some specified threshold. In contrast to this psychology-based approach, models of CPR management typically assume that agents respond rationally to available information in a manner described by evolutionary game theory. The two types of model can behave quite differently.

Large-scale CPR conservation efforts face psychosocial constraints and are also rationally motivated. They thus combine features of the above two model types. Agents will face material consequences in the long term and at least some people are aware that it is in their long-term self interest to try to mitigate these consequences through some short-term sacrifices. They might also realise that the collective action problem may be solved more quickly if they are willing to discuss with others. This means actively overcoming bounded confidence to engage with people around them even if they have very different views.

While long-term limiting (equilibrium) behaviour in both CPR problems and opinion dynamics has been widely studied~\cite{equil1,equil2}, the transient dynamics have received far less attention. 
Successful conservation, however, may depend not only on the long-time group behaviour but also on the rate of behavioural change once the problem is first recognised. A small number of opinion dynamics studies have specifically modelled transient dynamics~\cite{banisch,de} and provided evidence that this is important for understanding realistic systems of practical interest, which are often in such transient states.

In this work, we study the transient dynamics of collective action through behavioural change mediated by social interactions between agents who are also influenced by a broader sociopolitical environment. Being rationally motivated by the need for action, neighbours of different opinions overcome bounded confidence and are just as likely to interact with each other as neighbours of similar opinions. We also expose the agents \textit{en masse} to a global field that represents the combined effect of influences in the broader sociopolitical environment. We study the competition between these local and global influences, and how the speed of consensus formation emerges from different features of the model. Finally, we discuss implications of our findings for real-world management of large CPRs.

We study the spread of cooperation with a pro-conservation behavioural norm using a model that is similar to those used for opinion dynamics without bounded confidence \cite{castellano}. Each agent's level of cooperation is represented by a real number from -1 to 1, which respectively represent total defection and total cooperation, and we refer to agents at these extremes as defectors and cooperators accordingly. We adopt this nomenclature from evolutionary game theory even though our model more closely resembles an opinion dynamics model, to reflect the mixture of rationally motivated, strategic behaviour and psychosocial constraints in the norm-spreading dynamics that we aim to capture. Since we are concerned with building collective action from a state of inaction, our simulations begin with network states in which most agents defect.

We study dynamics on clustered scale-free networks (CSF).
Real social acquaintance networks are typically clustered scale-free  networks \cite{smallworld,smallworld2}. In such networks, topological path lengths between agents are typically short (even in the case of large and highly connected online social networks \cite{facebook}), giving rise to the name `small-world network'\cite{6degrees}. Although scale-free networks have been extensively studied \cite{barabasi}, frameworks for constructing and analysing clustered scale-free networks were developed only recently. In addition to their small-world properties, such networks have high connectivity between proximate nodes (i.e. topological clustering). However, we caution the reader that, except where referring to CSF networks directly, in our analysis we use `cluster' not in this topological sense but rather to refer to a group of agents who share a common cooperativity (i.e. strongly correlated states). We study collective dynamics across many simultaneously evolving network realizations and average the resulting network properties over the many realizations.  

\section{Results}
Figure~\ref{fig:standard} shows a typical network evolution, averaged over 500 network realizations. Parameters are standard (see Methods and Supplementary Material) except $\phi = 0.015$ for the external field. In this and similar plots, we present the average cooperativity and standard deviation. The various parameters are defined in the Methods section and summarised in Table~\ref{parameters}.
The standard deviation here is calculated by first calculating the standard deviation of the cooperativity in each individual network realization, and then averaging these.

\begin{figure*}%
    \centering
    \begin{subfigure}{0.45\textwidth}
    \includegraphics[width=\textwidth]{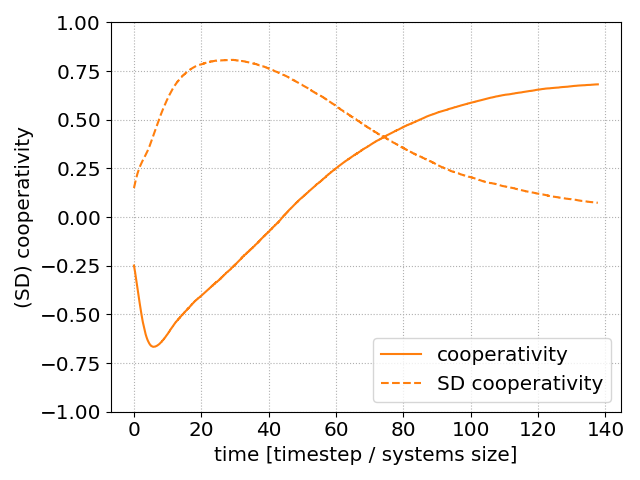}
    \caption{A Typical network evolution, here $\phi = 0.015$.}
    \label{fig:standard}
    \end{subfigure}
    ~
    \begin{subfigure}{0.45\textwidth}
    \includegraphics[width=\textwidth]{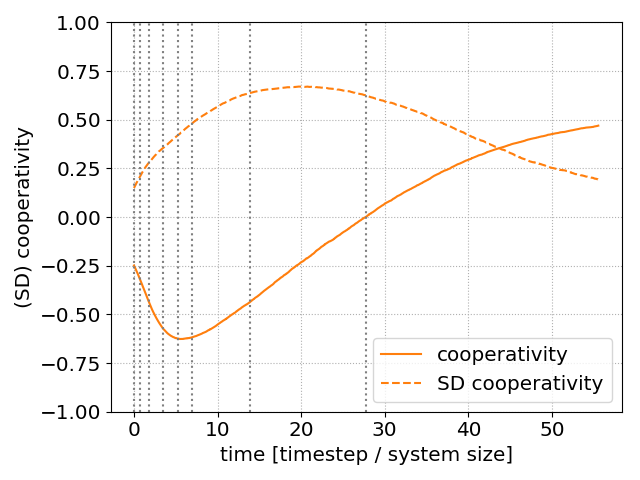}
    \caption{Grid network evolution with snapshots below.}
    \label{gridexample}
    \end{subfigure}
        \vskip\baselineskip
    \begin{subfigure}{0.23\textwidth}
    \includegraphics[width=\textwidth]{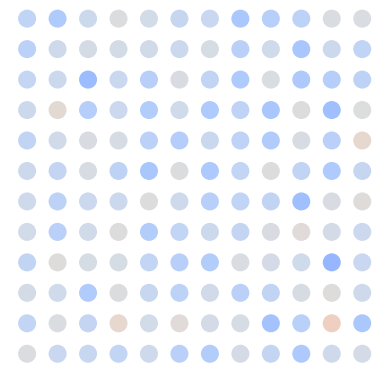}
    \caption{t = 0}
    \end{subfigure}
    ~
    \begin{subfigure}{0.23\textwidth}
    \includegraphics[width=\textwidth]{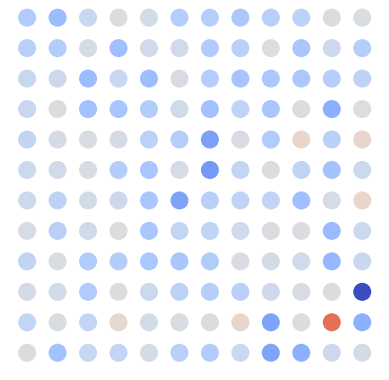}
    \caption{t = 0.7}
    \end{subfigure}
    ~
    \begin{subfigure}{0.23\textwidth}
    \includegraphics[width=\textwidth]{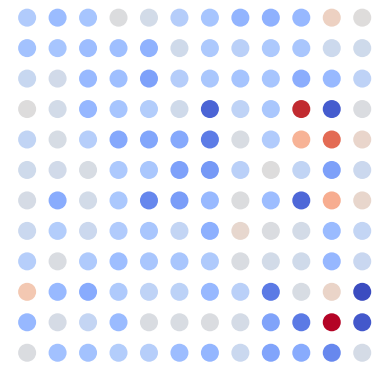}
    \caption{t = 1.7}
    \end{subfigure}
    ~
    \begin{subfigure}{0.23\textwidth}
    \includegraphics[width=\textwidth]{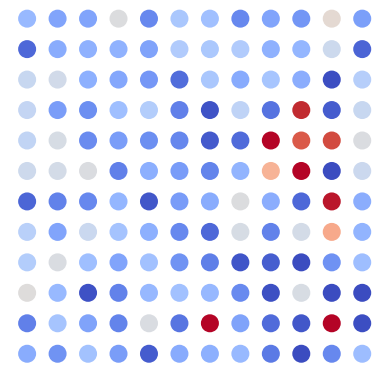}
    \caption{t = 3.5}
    \end{subfigure}
    \vskip\baselineskip
    \begin{subfigure}{0.23\textwidth}
    \includegraphics[width=\textwidth]{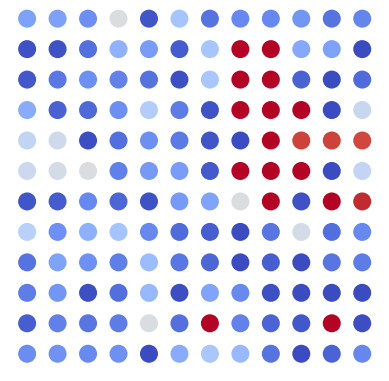}
    \caption{t = 5.2}
    \end{subfigure}
    ~
    \begin{subfigure}{0.23\textwidth}
    \includegraphics[width=\textwidth]{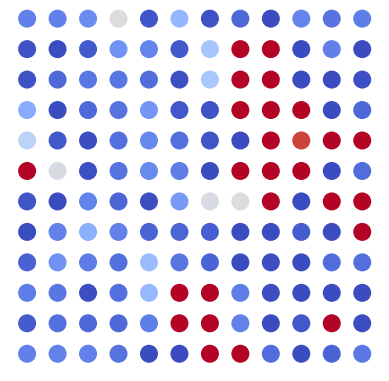}
    \caption{t = 7.0}
    \end{subfigure}
    ~
    \begin{subfigure}{0.23\textwidth}
    \includegraphics[width=\textwidth]{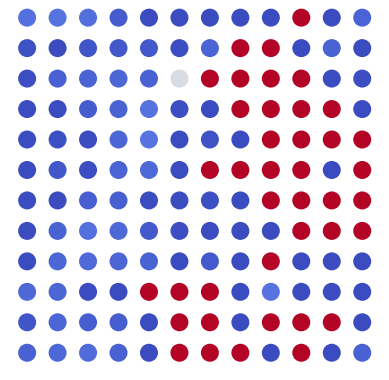}
    \caption{t = 13.9}
    \end{subfigure}
    ~
    \begin{subfigure}{0.23\textwidth}
    \includegraphics[width=\textwidth]{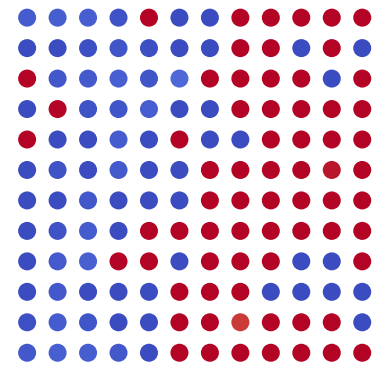}
    \caption{t = 27.8}
    \end{subfigure} 
    \vskip\baselineskip
    \begin{subfigure}{0.23\textwidth}
    \includegraphics[width=\textwidth]{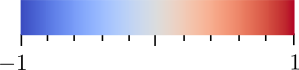}
    \end{subfigure}
    \caption{(a) A typical average time evolution of the cooperativity for conditions slightly favouring cooperation on a CSF network. At first, the short range but strong social pressure of the initially abundant defectors dominate the evolution. However, after some time, the long-range (global) but weak external field drives the networks slowly into cooperation. The standard deviation initially increases during cluster formation, but later decays away as the network homogenizes. (b) An example of how a grid network of small size evolves. Comparing (b) with grid snapshots (c-j) helps to understand how the clustering process evolves; the dashed lines in (b) indicate the times corresponding to (c-j). Moreover, comparing (b) with (a) shows the similarity of dynamics in the grid and CSF network cases. (c-j) A realization of a small grid network with periodic boundary conditions and an initial defector majority, but a weak field favouring cooperation. Blue indicates defective behaviour, red cooperative.  We see that, initially, agents with strong opinions efficiently convert undecided agents in their vicinity to their cause.  This is the cluster-formation regime of the dynamics. In this realization, one can observe a strongly cooperating (red) agent in the upper right, which appears at $t=1.7$.  This agent them forms a large cooperative cluster around it. In contrast, most of the rest of the network gradually becomes more defective due to the initial abundance of defectors. At $t=13.9$, the network is almost completely polarised, and by $t=27.8$, the cooperative cluster has begun to overcome the defective cluster by gradually dissolving its perimeter.}%
    \label{fig:examples}%
\end{figure*}

\subsection{Transient polarisation}



The dynamics show two characteristic time scales. Initially, we see a rapid decline in the average cooperativity and increase in its standard deviation. However, the dynamics soon become dominated by a second, slower process that gradually drives the network to a fully cooperative state. We can understand this better from simulations on smaller 13x13 grids with the same degree and periodic boundaries, shown in figure~\ref{fig:examples}, along with the evolution of the average cooperativity and standard deviation. Initially, neighbouring agents are uncorrelated. Agents that are initially strongly cooperative or defective spread their behaviours to their surroundings and strengthen their own behaviours. Correlated clusters emerge around the original agent and any interaction within the cluster will serve only to reinforce the behaviours, leading locally to an `echo chamber' and globally to a strongly clustered, polarised network. Once the network has become completely polarised, meaningful interactions take place only on the boundaries separating opposing clusters.

Comparing the snapshots in fig.~\ref{fig:examples} to the time evolution shows that this cluster formation regime corresponds to the first stage in figure \ref{fig:standard}. Since the network begins with a small majority of defectors, they on average spread their behaviours more effectively than cooperators.  Hence we see strong initial decrease in the average cooperativity, averaged over many networks.  Notably, such a polarised state is often the equilibrium to which bounded-confidence models with a sufficiently restrictive threshold converge.  In our system, however, it is only transient.  Once polarisation is complete and no or few neutral agents remain, a `tug-of-war' takes place between clusters, at their edges. In this competition, a sufficiently strong pro-conservation external field can tip the balance, making it more likely that one-on-one interactions will result in defectors converting to cooperators, rather than the converse. We see this in the second dynamical stage, in which the cooperativity slowly rises as defective clusters shrink.

\begin{figure}%
    \centering
    \begin{subfigure}{0.45\textwidth}
    \includegraphics[width=\textwidth]{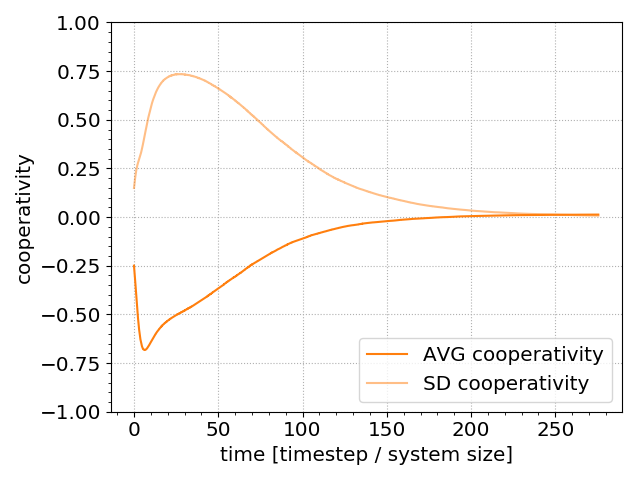}
    \caption{}
    \label{avgstate}
    \end{subfigure}
    ~
    \begin{subfigure}{0.45\textwidth}
    \includegraphics[width=\textwidth]{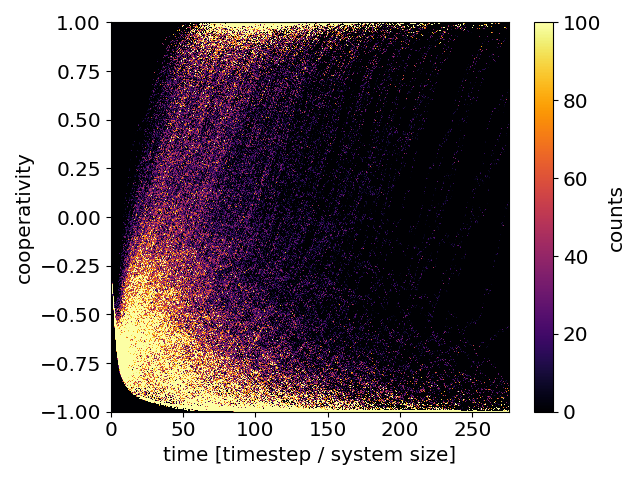}
    \caption{}
    \label{heatmap}
    \end{subfigure}
    \caption{Long-term evolution of a system close to a tipping point ($\phi = 0.0103$). (a) The average and standard deviations of the cooperativity converge to a value near zero, meaning that in roughly half of the realizations, the networks became completely cooperative, and in the other half, completely defective. This evolution is very sensitive to changes in parameter values around this point, as is a hallmark of tipping points.  (b) A heatmap corresponding to all the data for which the averages are shown in (a). The intensity corresponds to the density of networks having a specific average cooperativity at a specific time. Most networks converge to full cooperation or defection within the simulation time, reinforcing that this is a tipping point. }%
    \label{tipping}%
\end{figure}

Figure~\ref{fig:standard} shows that the point of maximum polarisation, where the standard deviation is maximal, does not coincide exactly with the point of minimal cooperation. In figures~\ref{fig:examples} c-j, we see that, at the cooperation minimum, there are regions that are not yet polarised but are responding to the pro-conservation external field.  The dynamical crossover from rapid polarisation to slow cluster growth/shrinkage thus happens somewhat after the minimum.  A consequence of this is that the average cooperativity is not bound to monotonically increase after the apparently minimal average cooperativity is reached. Rather, at the end of the cluster formation stage, in some cases, even though the average cooperativity has already started to increase slightly, the cooperators nevertheless are not abundant enough and the system converges to a fully defective state (Fig. \ref{heatmap}). 

We can understand the above results in more detail by considering the strengths of the two competing interactions.
More-or-less neutral agents are more susceptible than polarised agents to persuasion by others. Since neutral agents are more abundant in the initial network configuration, social interactions more strongly affect its early dynamics, which quickly converge to a polarised state. 
Once the system is strongly polarised, persuasion is possible only between agents of extreme, opposite levels of cooperativity at the cluster boundaries. In this situation, even a small contribution from the external field can tip the probability in one direction over the other.  Thus, the defector is slightly more likely to become a cooperator than the cooperator is to become a defector.  When this is the dominant dynamics, the network evolves more slowly, but the influence of the external field becomes apparent.


In some regions of the parameter space, we observe so-called social tipping points, around which the network's long-time behaviour depends sensitively on the parameters. Figure~\ref{tipping} shows a network evolution around such a tipping point. In panel (a), both the average and standard deviation of the cooperativity converge to roughly zero, implying that the individual network realizations converged to fully cooperative and fully defective states with roughly equal frequencies. Panel (b) shows a 2D histogram, binning the cooperativities of the individual network realizations that are averaged in (a). The colour scale is capped at 100 counts to enhance contrast. We see a clear division of the network realizations between the two attracting terminal states. The final outcome for each realization depends on its precise details.  The initial conditions were chosen such that fully cooperative and fully defective outcomes are both likely.  The external field here was $\phi = 0.0103$, and all other parameters standard.

\subsection{Convergence rates}
\begin{figure}
    \centering
    \includegraphics[width=1.0\linewidth]{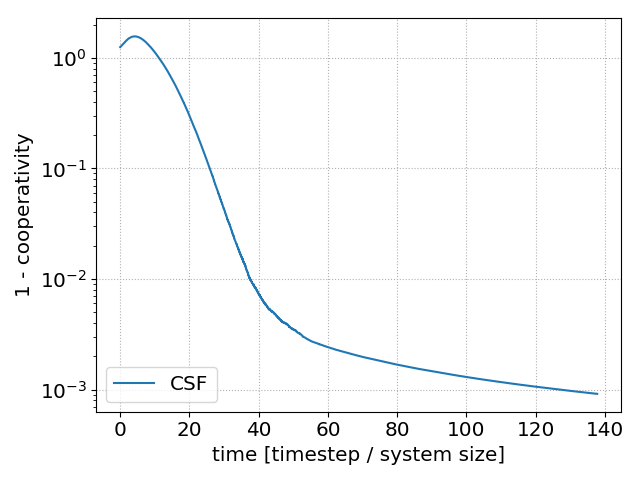}
    \caption{Logarithmic plot for a long range simulation of the average cooperativity in a standard clustered scale free network. We see that there is a critical slowing down at extremely homogeneous systems.}
    \label{logfig}
\end{figure}

Figure \ref{logfig} gives more insight into the process of convergence to a fully cooperative network by plotting the same transient dynamics as figure \ref{fig:standard} on a logarithmic scale. We see that the second, convergent dynamical stage in \ref{fig:standard} is itself separated into two stages. The first is an approximately exponential stage of convergence, as expected. In this second, slower, stage, it is likely the finite size of the remaining clusters that plays a role, in combination with fluctuations.

In real-world collective action problems, achieving the threshold needed for effective action is far more important than converting the last few hold-outs. Therefore it is the exponential stage of the convergence regime that interests us most. Since the length of this stage varies, we choose a simple and consistent way to evaluate the rate: taking the time derivative of the average cooperativity when the average cooperativity is close to 0, where the curve is certain to be exponential.

\subsubsection{Network connectivity}

\begin{figure*}%
    \centering
    \begin{subfigure}{0.45\textwidth}
    \includegraphics[width=\textwidth]{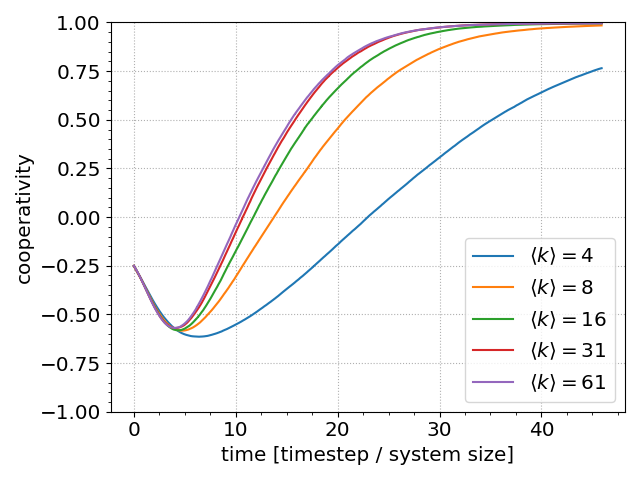}
    \caption{}
    \label{degree}
    \end{subfigure}
    ~
    \begin{subfigure}{0.45\textwidth}
    \includegraphics[width=\textwidth]{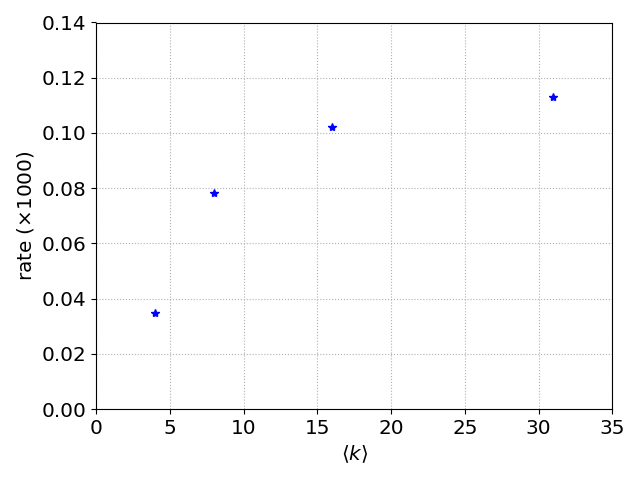}
    \caption{}
    \label{degrate}
    \end{subfigure}
    \label{degreefigs}%
    \vskip\baselineskip
    \begin{subfigure}{0.45\textwidth}
    \includegraphics[width=\textwidth]{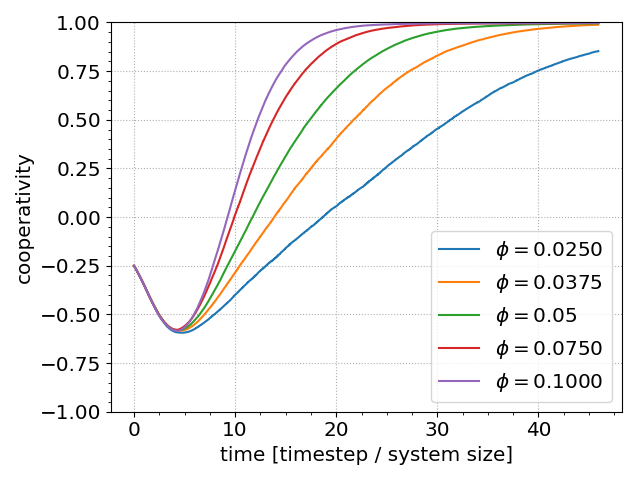}
    \caption{}
    \label{fieldfig}
    \end{subfigure}
    ~
    \begin{subfigure}{0.45\textwidth}
    \includegraphics[width=\textwidth]{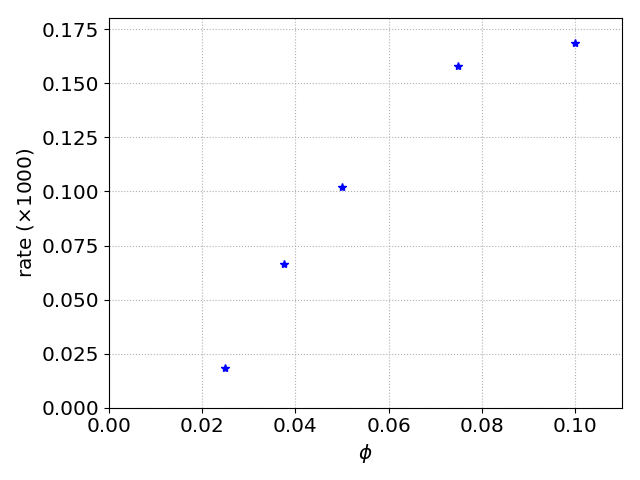}
    \caption{}
    \label{phirate}
    \end{subfigure}
    \caption{The effect of the connectivity and field on the convergence to the final state.  (a) and (b) the average cooperativity as a function of time for different network connectivities and corresponding rates. (c) and (d) the same for field strength.  Increasing the average connectivity in the network increases the convergence speeds, because each individual agent on average will have more ways of receiving the dominant message.  As expected from the model formulation and the analysis thereof, the strength of the external field is paramount to both the convergence speed as well as the fate of the networks.  We see that initially, improving the conditions of the networks yields a significant change in convergence rate. Whereas for already favourable conditions, there is a diminishing return on the gain in convergence rate.}%
    \label{phifigs}%
\end{figure*}

We investigate how the transient dynamics depend on network topology by comparing the convergence for different networks. Figure~\ref{degree} shows the average cooperativity evolution for clustered scale-free networks with different degrees and otherwise identical, standard parameters. We see that network connectivity is a crucial determinant of the time scale at which clusters are dissolved. In figure~\ref{degrate}, we show the convergence rate as a function of the average degree (obtained by changing the growth degree, $k_0$, to 4, 8, 16, 31, 61 respectively, see methods section). Once the network is sufficiently connected, the convergence rate saturates. This is most likely due to the finite size of the network. We can calculate the average number of next neighbours as the square of the average degree. Since our system contains around 1000 agents, if the average degree is 31 or 61, every agent is likely to be connected to a quarter of, or all (respectively), other agents, through its nearest neighbours. 

\subsubsection{External field}
We explore how the external field affects the convergence rate by first letting the network polarise (reach its minimum average cooperativity), and then changing the field. In this way we can discount any effects of the field on the initial polarisation stage as significant determinants of the convergence dynamics.

Figure \ref{fieldfig} shows the results. Apart from the delayed change to the external field, the parameters and simulation protocols were identical to earlier simulations. Each curve corresponds to 500 network realizations. As shown in figure~\ref{phirate}, the external field strength strongly affects the average cooperativity's convergence rate, though the response diminishes as field strength increases. 


\subsubsection{Influencers}

\begin{figure}
    \centering
    \includegraphics[width=1.0\linewidth]{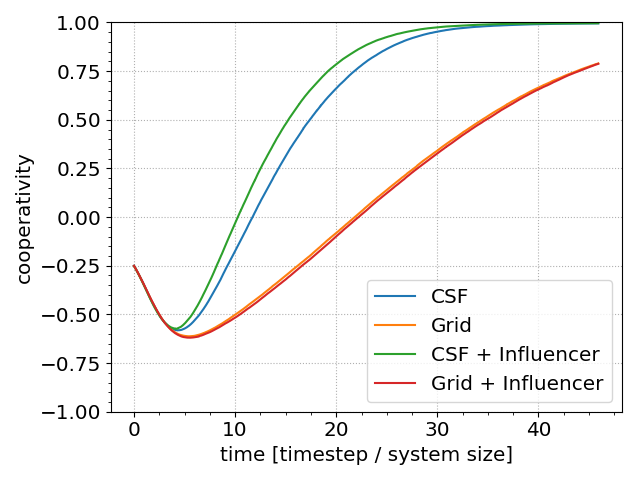}
    \caption{The average cooperativity as a function of time for different grid types with and without an \emph{influencer} in the network. An influencer is a highly connected (in this case the most connected) node with an unwavering behaviour. We can see that the key property of an influencer is the first of these attributes, as their effect in a grid network is almost negligible.}
    \label{incluencer}
\end{figure}

Over recent years, it has become clear that so-called \emph{influencers} can affect the spread of behaviours through social networks, particularly on social media. We represent such an influencer in our model as an agent who has both exceptionally high connectivity and strong, inflexible behaviour. Again, we first let the networks polarise and then begin our analysis by identifying the agent with the highest connectivity in the network and fixing this agent as an unyielding cooperator.
To verify that the subsequent dynamics reflect both this agent's behaviour and connectivity, we compare this with the addition of an influencer to a grid network in which all agents are equally well connected. The results are shown in figure \ref{incluencer}, again with standard parameters and 500 network realizations. Clearly, the presence of an influencer drives substantially faster convergence to the final state. 

\begin{figure}
    \centering
    \includegraphics[width=1.0\linewidth]{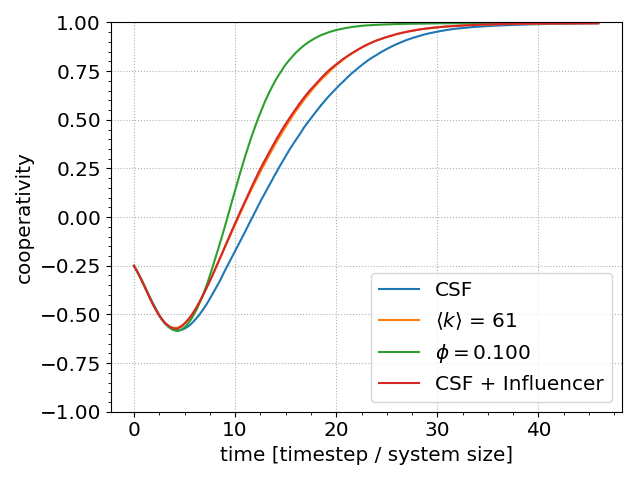}
    \caption{Summarizing cooperativity as a function of time for the various interventions suggested in this paper to increase the rate of convergence towards consensus. Data collected from previously presented figures. The yellow and red curves are overlaid because their impacts are almost identical.}
    \label{fig:comparison}
\end{figure}

Figure \ref{fig:comparison} compares the effects of our various interventions on the convergence rate. For brevity, we include only the strongest case observed for each intervention. Under the given assumptions, the external field most strongly affects the convergence rate. However, we emphasise that this result should be interpreted with caution, since we do not account for factors affecting the practical feasibility of each intervention in any real system. We also leave detailed exploration of interactions between interventions for future work. 

\section{Discussion}

In this study, we used an agent-based model to investigate the transient dynamics of collective action to conserve a large common-pool resource. This problem combines rationally motivated behaviour with the many unconscious processes intrinsic to social psychology. We used an opinion dynamics-type model, without bounded confidence, to capture this dual nature and studied the effects of network topology, an applied external field, and the presence of `influencers' on the rate at which the network population converged to a state of full cooperation with a pro-conservation behavioural norm. 

Different processes dominate the model on different time scales. Initially, formation of polarised clusters dominates but this gives way to a slower process of cluster dissolution due to social interactions at their perimeters, which depends strongly on cluster geometry. In larger networks, there is a possibility of larger clusters.  Since the number of agents inside clusters will grow faster with the cluster size than the number of agents at their perimeters, the convergence will likely be slower.

We found that the convergence rate is strongly affected by both network connectivity and external field strength, and that some parameter regions show tipping points in which the long-time behaviour depends sensitively on the parameters. A pro-conservation field gives cooperators the edge to slowly convert defectors at the cluster perimeters. We also found that a highly-connected and zealous influencer, who can convince others but is itself inflexible, can greatly speed up the convergence.

In the presence of any of these interventions, achieving collective action is essentially a process of dissolving defective clusters, which is fundamentally limited by network topology at the cluster perimeters. This result strongly agrees with recent work by Fotouhi et al \cite{cojoin}, who found that deliberately and carefully increasing connectivity between polarised clusters is a powerful mechanism for promoting cooperation. In our view, this speaks particularly to the power of now-ubiquitous social media echo chambers as inhibitors of adequate collective action on the extremely urgent global challenges of climate change and biodiversity loss. Our results suggest that replacing the mechanisms that generate and maintain such polarisation with others that instead promote constructive engagement between disagreeing groups could be a powerful promoter of large-scale collective action. This is an important subject for further modelling, including additional features of real social networks such as link rewiring, which can have a significant impact on formation and therefore also dissolution mechanisms for polarised social clusters~\cite{tuan,min2018}. 

Previous investigations have extensively and separately discussed social tipping points~\cite{scheffer2012}, clustering~\cite{hegselkrause,deffuant}, the role of connectivity \cite{cojoin}, as well as the impact of external parameters \cite{quatr}, all of which are represented in our model. Our results suggest that real-world efforts to promote collective action on large-scale problems can benefit from both improving network connectivity  \emph{and} providing a sociopolitical climate which favours cooperation. 
In this sense, we combine previous results into an overarching framework. We hope that this novel perspective will open new avenues into investigating the social dynamics that are critically important for addressing 21st-century global challenges.

Our results also suggest that convergence rate, rather than just long-time dynamics, are crucially important for real-world collective action. Much effort has been invested in trying to understand which parameters ensure a favorable final state of a social network. However, for urgent large scale common-pool resource problems, simply concluding that there exists a stable favorable solution that will be reached \emph{eventually} is in practice not sufficient to ensure a favourable outcome, since we also have to ensure that we will arrive there before it is too late.  Otherwise the commons may have already sustained significant, or even catastrophic and irreversible, damage. We would therefore argue that the complexity of large-scale common-pool resource management is such that considering a single intervention is insufficient for advising mitigation strategies. Ensuring favourable network connectivity would be moot if a favourable sociopolitical climate is lacking, and vice versa.

Finally a word on the scope of this research. Our model differs from the well-established bounded-confidence models that are conventionally used to model opinion dynamics, and which typically produce final states with clustering. This is founded on the fact that widespread consensus is generally not reached on the vast majority of questions about which people have opinions, such as which TV-show is best or which political party to vote for.
However, there are several large-scale societal issues on which opinion and/or behavioural homogenization is occurring, such as gender equality, democratization, and possibly the need for substantial climate change mitigation. All of these issues are characterized by the threat of long-term system-wide material consequences for inaction, giving agents a rational incentive towards cooperation. An interesting avenue for future research would be to introduce a parameter for the level of confidence bounding to our model and investigate the transition between the clustered final state typically seen in regular bounded confidence models and the final full consensus observed in the case of unbounded confidence analysed here. To put things into perspective, it should be noted that bounded confidence models with a large confidence interval can also produce a consensus end state. However, tuning which state becomes the end state is generally not a natural component of such models.

\section{Methods}

We study the spread of cooperation with a pro-conservation behavioural norm
using a formalism that is similar to other models of opinion dynamics without bounded confidence \cite{perc}. Each agent's level of cooperation ($b_i$) is represented by a real number from -1 to 1, which respectively represent total defection and total cooperation, $b_i \in [-1, 1]$.  Since we are concerned with building collective action from a state of inaction, our simulations begin with network states in which most agents defect.

The network evolution is simulated iteratively in 4 steps: 1. Randomly select an agent ($i$); 2. Randomly select a neighbour ($j$) of $i$; 3. Evaluate their interaction as described below. 4. Update $i$'s behaviour according to the outcome. This process is iterated until the network converges to a steady state. Updating $i$'s behaviour, rather than $j$'s, ensures that agents with more connections are more influential.

The change $\Delta b_i$ in $i$'s behaviour resulting from an interaction is given by\footnote{These equations differ slightly from the published version of the paper. One misprint was corrected and readability was improved.}
\begin{equation} \label{oldinteraction}
    \Delta b_i
    = |b_i - b_j| [p^+(1-b_i) - p^-(1+b_i)],
\end{equation}
where
$p^+$ and $p^-$ describe the weights of opinion changes in the positive and negative directions respectively.
Here $p^+ = f(x_{ij}, r, \xi)$ and $p^- = 1 - p^+$ with
\begin{equation} \label{rescaling}
    f(x,r,\xi) = 
    \begin{cases}
        0 & \text{if } x + \xi < -r\\ 
        \frac{1}{2r}(x + r - \xi) & \text{if } -r \leq x + \xi \leq r \\
        1 & \text{if } x + \xi > r, 
    \end{cases} 
\end{equation}
where $\xi$ is a uniform random noise term, $r$ is the width of the noise, and furthermore
\begin{equation} \label{interaction2}
        x_{ij} =  w_i b_i + w_{ij} b_j + \phi.
\end{equation}
 We can simplify equation \ref{oldinteraction} to
\begin{equation} \label{interaction}
    \Delta b_i
    = |b_i - b_j| [2 f( x_{ij}, r, \xi ) - b_i - 1] .
\end{equation} 

These equations are motivated as follows. The first is a continuous generalization of a discrete opinion exchange, as reviewed in \cite{castellano}. In such models, one would weigh the probability of an agent flipping its state when evaluating the agent interaction, this is the function of $w$ in equation \ref{oldinteraction}. Additionally, in this interaction we have added a prefactor that limits the degree to which two already agreeing agents can influence each other and, conversely, ensuring that an agent has to be presented with a new opinion in order to change its opinion.

Equation \ref{interaction2} describes the magnitude of the agent-to-agent interaction. The function $f(x_{ij}, r, \xi)$ captures the response of the affected agent, including the random environmental influences, which are in $r$.
Finally, equation \ref{interaction} is just a more compact version of equation \ref{oldinteraction} that is more useful for computations. 

We now define every parameter in the model (summarised in Table \ref{parameters}). The parameter, $\phi$ represents the influence of the sociopolitical environment, which we assume to affect all agents equally. Media influence has been variously studied in opinion dynamics models (e.g.~\cite{quatr,lizhu}), usually implemented to represent a specific mechanism or scenario. Since our focus is more general, $\phi$ is intended as a simple representation of the net influence of many factors including media messaging and regulations imposed by institutions.
A positive (negative) $\phi$ corresponds to a net positive (negative) influence on the agents' behaviour, making them more cooperative (defective). We use a small but positive external field, which represents a liberal society that is promoting cooperation on collective issues.

The agent's resistance to change is represented by $w_i$. Experimentation showed that our results do not qualitatively depend on $w_i$'s distribution across agents, so we assume them to be the same for all agents. The interaction strength between agents $i$ and $j$ is $w_{ij}$, which represents a combination of psycho-social factors such as relatedness, charisma, and cultural differences. We take this parameter to be normally distributed, since it reflects a combination of a large number of random influences. To accommodate the many highly variable instantaneous influences on social interactions, such as mood, location and medium of communication, we screen the interactions with a uniformly distributed stochastic noise term, $ -r \leq \xi \leq r $, where $r$ is the width of the distribution.

The factor, $|b_i - b_j|$ in eq.~(\ref{interaction}) is introduced to ensure that the final opinion of an agent after an interaction lies inside or at least near the range of opinions of the interacting agents. This reflects the fact that agents can only change to other opinions that are actually presented to them.
Consequently, meaningful changes in agents' opinions primarily occur in exchanges with agents not already sharing their opinion. Conversely, any interaction between two already agreeing agents will only work preserve the agents' present opinions.

Since we are concerned with building collective action from a state of inaction, our simulations begin with network states in which most agents defect.
Except where specified otherwise, they begin from an uncorrelated Gaussian distribution of agent behaviours with mean $-0.25$ and standard deviation $= 0.15$. 

We use the Holme-Kim algorithm \cite{holmekim} to construct our clustered scale-free networks. We also analyse simple grid networks for pedagogical purposes.
Unless otherwise stated, all of our networks consist of $33^2 = 1089$ nodes, with average degree $\langle k \rangle = 16$. The average degree is obtained as a combination of the degree of the growth nodes added by the Holme-Kim algorithm ($k_0=8$), as well as the accompanying clustering parameter ($k_t = 3.5$).
We further use external field strength $\phi = 0.05$, agent self weight $w_i = 0.6$ and an inter-agent weight normal distribution with mean $0.5$ and standard deviation $0.15$. Furthermore, the uniform interaction noise term $\xi$ is bounded by $r = 0.1$. Under these parameters, which we call standard, the network always converges to a fully cooperative state, allowing us to focus our investigation on the dynamics of this process.  All of the above parameters are summarized in table \ref{parameters}.

The stubbornness and friendship parameter values were loosely chosen in the mid range because with these parameters the model behaves in a way that is qualitatively similar to what is observed in the real world: people can influence each other but are generally reluctant to change their opinions, and this varies widely due to many parameters~\cite{mallinson}. We propose that refining this aspect of the model is an interesting avenue for future work, based on data from ongoing empirical work on opinion change.   

\begin{table*} 
\centering
\setlength{\extrarowheight}{7pt}
\begin{tabular}{llr} 
    \toprule
    Parameter/Variable                       & Represents                                    &    Value  \\
    \midrule
    Agent behaviour ($b$)           & The strategy (cooperative or defecting)   & - \\[-7pt]
                                    & that an agent operates                    & \\
    External field ($\phi$)         & Global factors:                               & 0.05      \\[-7pt]
                                    & Policy regulation, mass media, morals, etc    &           \\
    Self weight ($w_i$)             & Resistance to influence:                      & 0.6       \\[-7pt]
                                    & Stubbornness, ideological commitment, etc     &           \\      
    Agent-agent weight ($w_{ij}$)   & Susceptibility                                & 0.5       \\[-7pt]
                                    & Friendship, animosity, charisma, etc          &           \\
    Agent-agent weight STD          & Susceptibility variance                       & 0.15      \\
    Initial cooperation ratio       & Initial average cooperativity                 & $-0.25$   \\
    Initial cooperation STD         & Variance in initial cooperativity             & 0.15      \\
    Randomness ($r$)                & Noise intensity in agent interaction          & 0.1       \\
    Growth degree                   & Degree of nodes added during network growth  & 8      \\[-7pt]
                                    & as introduced by the Holme-Kim algorithm     & \\
    Clustering                      & Relative level of clustering of networks     & 0.5     \\[-7pt]
                                    & as introduced by the Holme-Kim algorithm     & \\
        \bottomrule
\end{tabular}
\caption{Summarizing all model parameters and variables to provide an overview of the model's components. Here the reference value of each parameter is given, which is used throughout this paper unless otherwise explicitly stated.}
\label{parameters}
\end{table*}

We study the collective dynamics across multiple simultaneously evolving networks. By obtaining data about the average cooperativity and standard deviation of cooperativty of agents in the networks, we can learn about what factors impact the network evolution and how. Throughout this paper, whenever we refer to the \emph{standard deviation}, we are referring to the average of standard deviations across the individual networks. The average state contains quantitative information about the evolution of the system, such as in which direction it is evolving, and at which rate. Combining this with the standard deviation, and prior knowledge about the network topology, we can draw qualitative conclusions about how the network is evolving and what mechanisms drive this evolution.

All simulations were run using \verb|Python 3.7.2| with the \verb|Networkx 2.3| (for network support) and \verb|Community 1.0.0b1| (for the Louvain algorithm \cite{louvain}) packages. Native parallelization routines were provided by the \verb|multiprocessing| package.

Wherever we refer to the convergence rate in quantitative terms, we use the derivative at average cooperativity zero as an indication of the this rate.
The derivative itself was found by linear regression around the data point closest to the root of the curve, we used a 10 point regression to safely eliminate numerical instabilities. 

For a more thorough explanation of the model parameter values, as well as the model's dependence on these, we refer to the Supplementary Material.

\section*{ACKNOWLEDGMENTS}
This work was supported by Swedish Research Council (Vetenskapsrådet) grant 2015-04962, Austrian Science Fund (FWF) grant P29252, Austrian Research Agency (FFG) grant 882184 and European Research Council (ERC) grant 682472-MUSES.
We acknowledge Signe Savén for ideas that inspired this research, and Tuan Pham for enlightening discussions. We also extend our gratitude to Effective Altruism for promoting collective action on important issues.









\twocolumn[{\LARGE \bf Supplementary material: Dynamics of collective action to conserve a large common-pool resource}\vspace{20pt}]







\section{Parameter dependence}

In order to ensure the fidelity of the model across the parameter space, we here present sweeps over the model parameters. In summary, when the parameters are varied, the model behaves as expected based on the findings presented in our main text. We comment below on the parameter dependence in the order of their appearance in figure \ref{fig:parameters}. Here we used 200 network realizations per simulation.

\subsection{External field ($\phi$), fig~\ref{fig:parameters}a}

Here we expand the interval over which the external field is varied. We see how an extremely strong positive external field (0.25) eliminates the initial decrease of the average state. This demonstrates that the external field is so strong that every single interaction increases the cooperativity. Beyond this point, increasing the external field further produces no qualitative changes. However, we see an even faster convergence to absolute cooperation.

At the other extreme, of low external field (0.01), we observe that the average state barely increases with time. The previous initial dip now resembles a drop in the average state, to a persistently low value. Decreasing the external field even further, we see a rapid drop to complete defection, from which the system never recovers. Its monotonic decline closely mirrors the increase we see for extremely positive external fields.

\subsection{Stubbornness ($w_i$), fig~\ref{fig:parameters}b}

Whereas one might expect a monotonic dependence of the cooperativity on the stubbornness, we instead see an optimal value at or near 0.6. We understand this intuitively as follows. At high stubbornness, the entire population is resistant to change and this slows the convergence.  At low stubbornness, agents are extremely susceptible to the opinion of any agent they interact with and the convergence is dominated by the random selection of interaction partners, rather than by the dynamics of the interaction, resulting in a very small net increase.

\subsection{Initial state, fig~\ref{fig:parameters}c}

The total time to reach a certain proximity near the final state depends on the initial state, because for lower initial cooperativity, there are more defectors to convince. For unfavourable enough conditions the abundance of defectors will be so large that they completely overtake the cooperators, and the networks will on average go fully defecting.  The red line for -0.5 levels off at a value not equal to 1 or -1.  This is because we are plotting an average over many realisations of the network, and in this case the final result (1 or -1) is so sensitive to the initial conditions and realisation of the interaction that some networks become fully cooperative and some become fully defective.  This is a finite-size effect.

\subsection{Initial state standard deviation, fig~\ref{fig:parameters}d}

We see that a larger variance in the initial state distribution corresponds to faster convergence. This can be understood from the grid figures in the main text. Some slightly cooperative agents must be present for cooperative clusters to form. The agent interaction (see Methods) is constructed such that agents on average can switch opinions only to other opinions that are presented to them. So, if no cooperators are initially present, none can appear subsequently either.

\subsection{Friendship ($w_{ij}$), fig~\ref{fig:parameters}e}

As friendship is implemented in the agent interaction almost identically to stubbornness, it is not surprising that the effects of changing the two are qualitatively similar. Friendship acts conversely to stubbornness: weak friendship makes it easier for an agent to retain their state and strong friendship makes the agent susceptible to changing its state.

\subsection{Friendship standard deviation, fig~\ref{fig:parameters}f}

Since we have chosen a favourable friendship value, giving it a large standard deviation introduces more nodes that either have very strong or weak friendships. This slows down the convergence, as described in the previous subsection.

\begin{figure*}%
    \centering
    \begin{subfigure}{0.40\textwidth}
    \includegraphics[width=\textwidth]{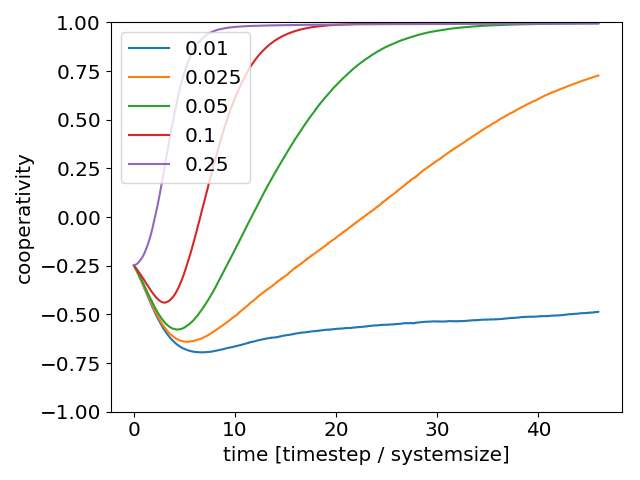}
    \caption{External field}
    \label{fig:standard}
    \end{subfigure}
    ~
    \begin{subfigure}{0.40\textwidth}
    \includegraphics[width=\textwidth]{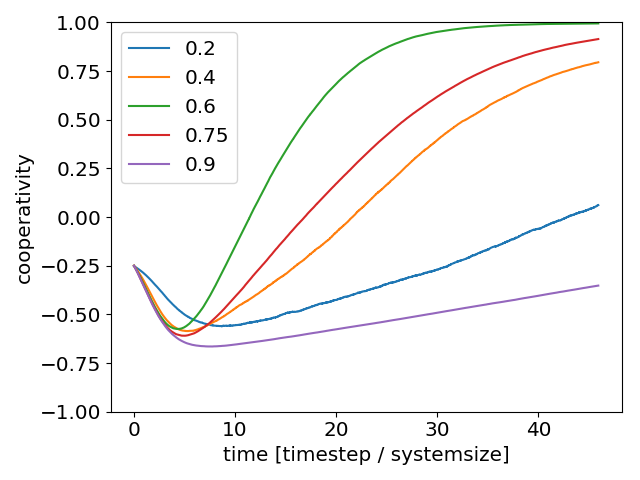}
    \caption{Stubbornness.}
    \label{gridexample}
    \end{subfigure}
    \vskip\baselineskip
    \begin{subfigure}{0.4\textwidth}
    \includegraphics[width=\textwidth]{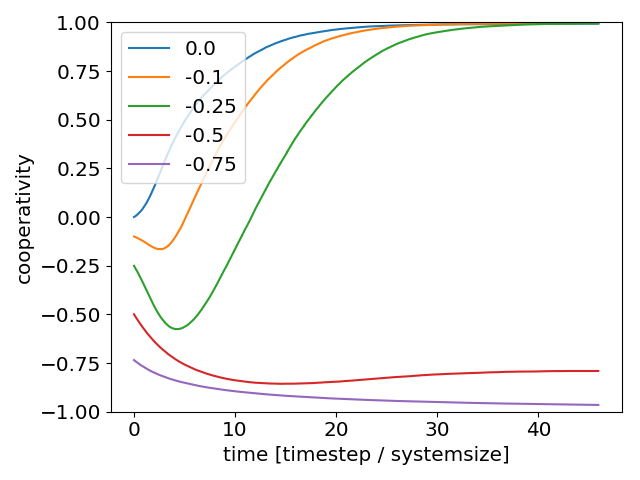}
    \caption{Initial cooperativity.}
    \label{fig:standard}
    \end{subfigure}
    ~
    \begin{subfigure}{0.4\textwidth}
    \includegraphics[width=\textwidth]{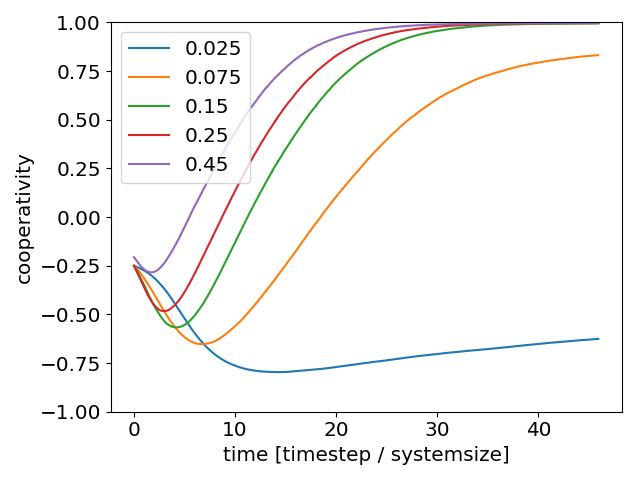}
    \caption{Initial cooperativity standard deviation.}
    \label{gridexample}
    \end{subfigure}
    \vskip\baselineskip
    \begin{subfigure}{0.4\textwidth}
    \includegraphics[width=\textwidth]{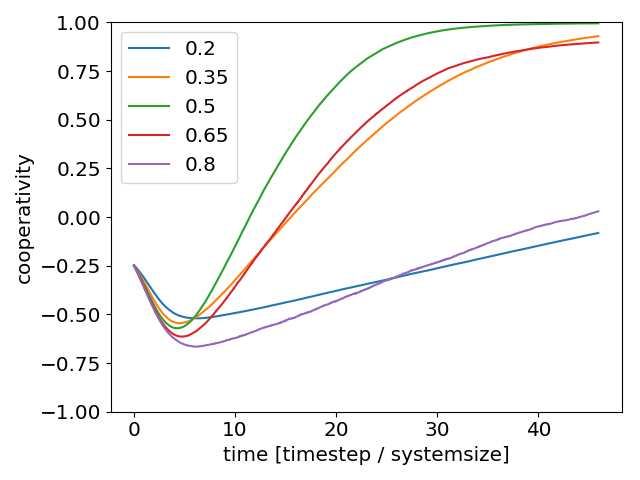}
    \caption{Friendship.}
    \label{fig:standard}
    \end{subfigure}
    ~
    \begin{subfigure}{0.4\textwidth}
    \includegraphics[width=\textwidth]{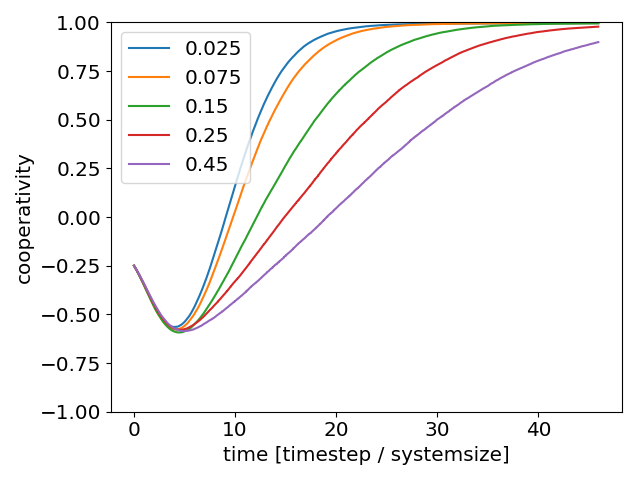}
    \caption{Friendship standard deviation.}
    \label{gridexample}
    \end{subfigure}
    \vskip\baselineskip
    \begin{subfigure}{0.4\textwidth}
    \includegraphics[width=\textwidth]{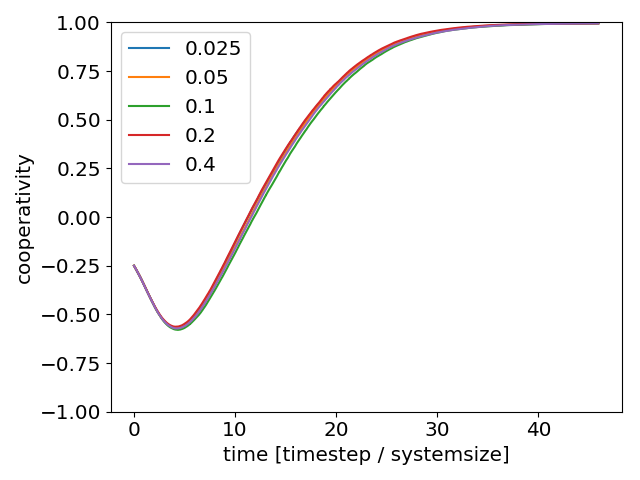}
    \caption{Randomness.}
    \label{fig:standard}
    \end{subfigure}
    \caption{Summarizing parameter dependence of the model, c.f. table I in the methods section of the main text.}%
    \label{fig:parameters}%
\end{figure*}

\subsection{Randomness, fig~\ref{fig:parameters}g}

The randomness parameter in the interaction does not strongly influence the convergence rate. Each individual realisation can vary more when the randomness is large, but the average convergence does not change.



\end{document}